\documentclass[%
aip,
jmp,%
amsmath,amssymb,
reprint,%
onecolumn,
nofootinbib
]{revtex4-1}

\usepackage[utf8]{inputenc}
\usepackage{graphicx}

\begin{document}

%\preprint{AIP/123-QED}

\title{Observing off-resonance motion of nanomechanical resonators as modal superposition}

%\author{Joshoua Condicion Esmenda $^*$}
\author{Joshoua Condicion Esmenda}
\thanks{These two authors contributed equally }
 \email{jesmenda@gate.sinica.edu.tw.}
 \affiliation{Department of Engineering and System Science, National Tsing Hua University, Hsinchu 30013, Taiwan}
 \affiliation{Nano-Science and Technology Program, Taiwan International Graduate Program,Academia Sinica, Taipei 11529, Taiwan}
 \affiliation{Institute of Physics, Academia Sinica, Taipei 11529, Taiwan}
 
%\author{Myrron Albert Callera Aguila $^*$}
\author{Myrron Albert Callera Aguila}
\thanks{These two authors contributed equally }
 \email{maguila@gate.sinica.edu.tw}
 \affiliation{Department of Engineering and System Science, National Tsing Hua University, Hsinchu 30013, Taiwan}
 \affiliation{Nano-Science and Technology Program, Taiwan International Graduate Program,Academia Sinica, Taipei 11529, Taiwan}
 \affiliation{Institute of Physics, Academia Sinica, Taipei 11529, Taiwan}

\author{Jyh-Yang Wang}
 \affiliation{Institute of Physics, Academia Sinica, Taipei 11529, Taiwan}
 
 \author{Teik-Hui Lee}
 \affiliation{Institute of Physics, Academia Sinica, Taipei 11529, Taiwan}
 
\author{Chi-Yuan Yang}
 \affiliation{Institute of Physics, Academia Sinica, Taipei 11529, Taiwan}
 \affiliation{Department of Physics, National Taiwan University, Taipei 10617, Taiwan}

\author{Kung-Hsuan Lin}
 \affiliation{Institute of Physics, Academia Sinica, Taipei 11529, Taiwan}
 
\author{Kuei-Shu Chang-Liao}
 \affiliation{Department of Engineering and System Science, National Tsing Hua University, Hsinchu 30013, Taiwan}

\author{Nadav Katz}
\affiliation{Racah Institute of Physics, Hebrew University, Jerusalem, 91904 Israel}

\author{Sergey Kafanov}
\affiliation{Department of Physics, Lancaster University, LA1 4YB, Lancaster, United Kingdom}

\author{Yuri Pashkin}
\affiliation{Department of Physics, Lancaster University, LA1 4YB, Lancaster, United Kingdom}

\author{Chii-Dong Chen}
 \affiliation{Institute of Physics, Academia Sinica, Taipei 11529, Taiwan}

\begin{abstract}
Observation of resonance modes is the most straightforward way of studying mechanical oscillations because these modes have maximum response to stimuli. However, a deeper understanding of mechanical motion could be obtained by also looking at modal responses at frequencies in between resonances. In this paper, we present visualisation of the modal response shapes for a mechanical drum driven off-resonance. Furthermore, we describe these transitions through superposition of resonance modes. 
\end{abstract}

%Use showkeys class option if keyword

\maketitle

\begin{quotation}
Observation of resonance modes is the most straightforward way of studying mechanical oscillations because these modes have maximum response to stimuli. However, a deeper understanding of mechanical motion could be obtained by also looking at modal responses at frequencies in between resonances. A common way to do this is to force a mechanical object into oscillations and study its off-resonance behaviour. In this paper, we present visualisation of the modal response shapes for a mechanical drum driven off resonance. By using the frequency modal analysis, we describe these shapes as a superposition of resonance modes. We find that the spatial distribution of the oscillating component of the driving force affects the modal weight or participation. Moreover, we are able to infer the asymmetry of the drum by studying the dependence of the resonance modes shapes on the frequency of the driving force. Our results highlight that dynamic responses of any mechanical system are mixtures of their resonance modes with various modal weights, further giving credence to the universality of this phenomenon.
\end{quotation}

\section{Introduction}
Mechanical vibrations of a structure, regardless of its geometry and type of material, can be fundamentally described by a combination of its natural resonance modes, called eigenmodes,  and a form of the driving force \cite{Verboven2002}. A resonance mode is defined as a pattern of motion in which the system or its part moves in a periodic oscillation with a maximal amplitude at a characteristic frequency. The spatial distribution of motion is determined by the inertial and elastic properties of the oscillating body, its shape, and the boundary conditions imposed on it. If either the material properties or the boundary conditions of the structure change, the resonance modes change accordingly\cite{Silvan2001}. While the resonance modes are characteristic of the inherent properties of the structure, their visualisation requires the application of a driving force, which provides an external stimulus by which one can excite these resonance modes. An example of such force is the time-varying electrostatic force between the two plates of a parallel-plate capacitor when an alternating voltage is applied across the plates\cite{Bunch2007}. Such parallel-plate geometry is often used in the experiments to study resonant properties of one of the plates, which is made flexible and can oscillate under the applied force. Regardless of the type of the driving force, a frequency response spectrum of a mechanical object is typically used to observe maximum response amplitudes at the natural frequencies. However, whereas this analysis allows the observation of resonant frequencies, it lacks spatiality that is necessary to observe the modal shapes. Several experimental studies on mechanical devices have addressed the issue of spatial resolution but focused only on the resonant modal shapes\cite{Zenghui2015,Davidovikj2015,Kim2018,wang2014}. Yet the question remains about the shape of these modal responses at off-resonance frequencies. Observing the off-resonance motion may provide insights into how the modal shapes transition from one to another.
%New Paragraph
\\
\\Modal analysis has been an important engineering tool for the past several decades \cite{Zhi2001}. In determining, improving, and optimizing dynamic characteristics of structures and components, it has found various applications in civil and nuclear engineering, aviation, automotive and space  industries, acoustic instruments production and biomechanical studies. As modal analysis is used in the numerical analysis of finite element methods, the presence of damping and nonlinearity requires experimental techniques to complement the computer model. Experimental modal testing involves both the application of a driving force and detection of vibration responses at various locations of the structures\cite{Schwarz2000}. While in most cases, modal analysis is performed for macrostructures, the concept behind the theory is not limited by the size. In fact, it could also be applied to small mechanical structures, like nanomechanical resonators (NMRs). Two-dimensional (2D) NMRs, in particular, found their niche as sensitive tools for measurements of various properties such as electrical conductance, thermal conductance, mass, radiation power, and many more \cite{Silvan2001,Bagci2014,DeAlba2016,Bunch2007}. Unique mechanical properties of 2D materials, which include low mass, high flexibility, and high tensile strength among others, allow them to have a large amplitude of flexural motion, making the NMR a viable object for coupling to other systems\cite{Andrews2014}. However, most of the previous studies of 2D NMRs lack spatiality. One of the recent studies that involves spatial resolution of mechanical response shows the inherent robustness of new multimode resonances from structural non-idealities\cite{Zenghui2015}. Another work improved upon this by visualising the resonant modes with local phase sensitivity to show that the nodal lines get deformed by small imperfections in the drumhead\cite{Davidovikj2015}. It was also shown by studying the spatially resolved resonant modes of two-dimensional bimorphs that the latter behave like anisotropically tensioned membranes\cite{Kim2018}. All of these studies used spatiality to identify the mode shape at resonance frequencies. However, they did not attempt to observe the resonator response and mode shapes at off-resonance frequencies.
%New Paragraph
\\
\\In this paper, we explore the effect of the driving force on the off-resonance motion. We investigate the response shapes, for driving frequencies both at resonance and off resonance, of the two-dimensional mechanical plate drums made from niobium diselenide (NbSe$_2$) flakes using a Fabry-Perot laser interferometer\footnote{The choice of NbSe$_2$ as the drumhead, while might be arbitrary for this work, could be beneficial for future low temperature studies as it's one of the 2D materials that are superconducting\cite{sengupta2010}.}. By observing the off-resonance response shapes, we see transitions from one resonance mode to another. Using the frequency modal analysis, we determine how the resonance modes of the system participate in vibrations, with some being more prominent than the others, as the driving frequency changes. A consequence of this is the demonstration of how the resonant modes depend on the oscillating component of the driving force spatial distribution. Also through this analysis, we infer the inherent asymmetry of the system from the resonant modes distribution with respect to the driving frequency.

\section{Results}
\subsection{\label{sec:level2}Description and characterization of the device}
NbSe$_2$ mechanical drums are fabricated through the transfer of exfoliated flakes onto a silicon chip with pre-patterned gold electrodes and AR-P (AllResist Positive) resist. The latter acts as a spacer for the drum (details of the fabrication are described in the Methods section and Supplementary Information). Figure 1a shows an optical image of a sample containing a circular drum (device A), and an elliptical drum (device B). Figure 1b shows the schematic cross-section of the chip along the white dashed line shown in Figure 1a, where a NbSe$_2$ flake is placed on top of the spacer. The circular and elliptical openings in the spacer define the drumhead locations, and a large rectangular opening on the right-hand side serves as the contact area for the flake with the gold electrode underneath. The motion of the mechanical drum is then detected using a laser interferometry technique as illustrated in Fig. 1c. A continuous-wave green laser beam (532 nm) is focused on the drumhead and the intensity modulation caused by the interfering reflections from the semi-transparent flake, like NbSe$_2$, and gold electrode underneath is captured by an avalanch photodetector. A scanning mirror is used to move the laser spot across the drumhead thereby making it possible to spatially resolve the recorded driven response. Through this method one can see how the response shapes look and also visually differentiate and ascertain the driven resonant modes, especially the higher ones.
%New Paragraph
\\
\\The driving of the mechanical drums is done using the electromotive scheme by supplying a voltage from a function generator with an oscillating component ($V_{ac}$) with a frequency close to the mechanical resonance frequency of the drum, and a dc component ($V_{dc}$). The dc component of the applied voltage allows amplification of the ac component, which improves the signal-to-noise ratio without the need of strong drive that might result in the non-linear regime of the resonator\cite{LeeJ2018}. A lock-in amplifier records the signal from the avalanch photodetector with a reference signal coming from the function generator. Figure 1d shows the magnitude and phase response at the fundamental resonance mode, which corresponds to a simple harmonic oscillator. Figure 1e is a three-dimensional plot of the spatially resolved frequency response mapping of the fundamental mode. The setup maps the frequency dependence of the mechanical displacement, $Z(x,y,\omega_d )$, where $x$ and $y$ are the mapping coordinates with the centre of the drum as the origin, to the corresponding photodetector voltage signal $V(x,y,\omega_d )$, where $\omega_d$ is the driving frequency. When $\omega_d$ is chosen to be the modal resonance frequency $\omega_{mn}$, where $m$ and $n$ refer to the number of nodal diameters and the number of nodal circles, respectively, for a clamped circular plate (the mode notation used is $(m,n)$), $V(x,y,\omega_d )$ reflects the resonance mode shape $Z_{mn} (x,y)$. When $\omega_d$ is off-resonance, $Z(x,y,\omega_d )$ is the shape of the off-resonance response. The details of conversion of $V(x,y,\omega_d )$ into $Z(x,y,\omega_d )$ are discussed in a separate article \cite{Myrron2020}.

\begin{figure}
    \centering
    \includegraphics[width=1.0\columnwidth]{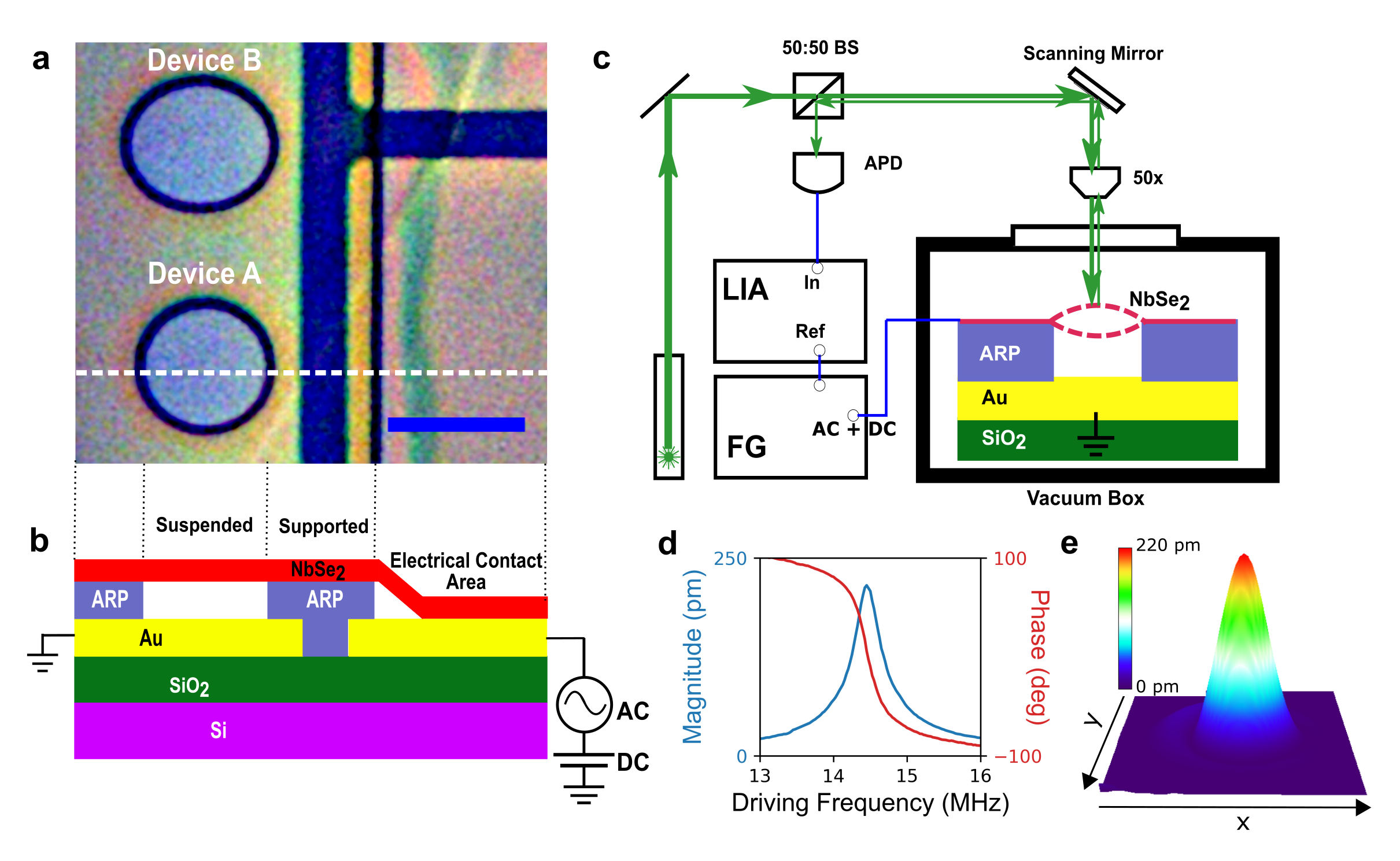}
    \caption{Characterisation of NbSe$_2$ mechanical drums using laser interferometry. (a) Optical image of the two mechanical drums (scale bar is 7 $\mu$m). (b) A schematic cross-section of the mechanical drums device. NbSe$_2$ flake (red) is placed on top of a patterned 295 nm thick AR-P electron beam resist (blue), which serves as a spacer. The flake is estimated to be about 55 nm thick using FEM simulations. There are two types of drumhead patterns, namely, a circle (device A)  with a diameter of 7 um, and an ellipse (device B) with minor and major axes lengths of 7 and 8 um, respectively. In a couple of microns away from the drumhead patterns, there is a larger rectangular opening on the AR-P resist spacer, where the flake is made to have electrical contacts with the gold bottom electrodes (in yellow). (c) The sample is contained inside a vacuum box with a pressure of about $10^{-7}$ mbar. Detection of motion of the mechanical drums is done by focusing a continuous wave green (532 nm) laser on the drumhead and recording the interfering reflections from the flake and the gold electrode underneath using an avalanch photodetector (APD). A scanning mirror is used to move the laser spot across the drumhead to record spatially resolved responses of the drums. The beam splitter (BS) is used to direct the reflected signal to the APD. A lock-in amplifier (LIA) then records the data from the APD while referenced by a function generator (FG). The drums are driven with a combination of AC and DC from a function generator. (d) Frequency response (magnitude and phase) of the fundamental resonance mode of device A. (e) Corresponding three dimensional plot at resonance of the magnitude as a function of x and y of device A. The z units are in pm and the spatial scanning range is about 12 $\mu$m x 12 $\mu$m. The magnitude of the driving signal is 250 mVpp with a DC of 4 V.}
    \label{fig:Figure_1}
\end{figure}

\subsection{\label{sec:level2}Spatial response mapping}
In order to visualise the response shapes, a system response to the drive at various frequencies is first measured for both devices at certain spots of their respective drumheads. From the knowledge of the theoretical mode shapes of a circular plate\cite{Timoshenko1928,Leissa1969,Mehta2009}, we could predict the frequencies of the resonance modes that are being driven (shown in the Supplementary Information). We also know that the lowest six modes are (0,1), (1,1)a, (1,1)b, (2,1)a, (2,1)b, and (0,2) in order of increasing frequency value, with the assumption of asymmetry. The a and b in the notation are used to designate the modal pairs with the same number of nodes but different orientation.
\\\indent
First, we investigate the response of device A. In Fig. 2a, two frequency response spectra are taken at different spots of the drumhead; one at the centre and the other about halfway to the edge. The spectrum taken at the centre has two peaks. This is indicative of the resonant modes that do not have nodal diameters passing through the centre. This description fits the fundamental mode (0,1), whose spatially resolved mapping is shown already in Fig. 1e, and the (0,2) mode having a higher frequency. Now, to find the other resonance modes, we look at the spectrum taken off the centre. It is important to note here that it may take several trials to carefully choose a spot that shows resonant peaks as each drum has its own nodal orientation. In the second spectrum, we identify four peaks, with frequencies labeled $f_0$ – $f_3$, and with $f_0$ having a $Q$ factor of 26, and the higher modes having $Q$ factors of 15, 18, and 21 for $f_1$, $f_2$, and $f_3$, respectively. The values of $f_0$ and $f_3$ are very close to the peak frequency values identified in the first spectrum, which indicates that the same modes, (0,1) and (0,2), are excited. This is indeed true as demonstrated in Fig. 2e, where the mapping at $f_3$ is shown and it resembles the (0,2) mode shape. The slight frequency shift between the two spectra is presumably due to laser heating causing a temperature gradient across the drumhead, which in turn causes the material to expand, inducing changes in the tension\cite{davidovikj2018,wang2014}. We do the same for $f_1$ and $f_2$. Figures 2b and 2c present the spatial mapping at frequency $f_1$ and the shape resembles (1,1) mode. In Fig. 2d, the spatial mapping at frequency $f_2$ is shown and the shape resembles the (2,1) mode. However, the corresponding modal pairs of the $f_1$ and $f_2$ were difficult to find. A possible explanation could be their non-favourability given the intrinsic asymmetry of the system. Now that we have mapped the response shapes at the resonances of the drum, we proceed to map the response shapes at frequencies between these resonances.

\begin{figure}
    \centering
    \includegraphics[width=1.0\columnwidth]{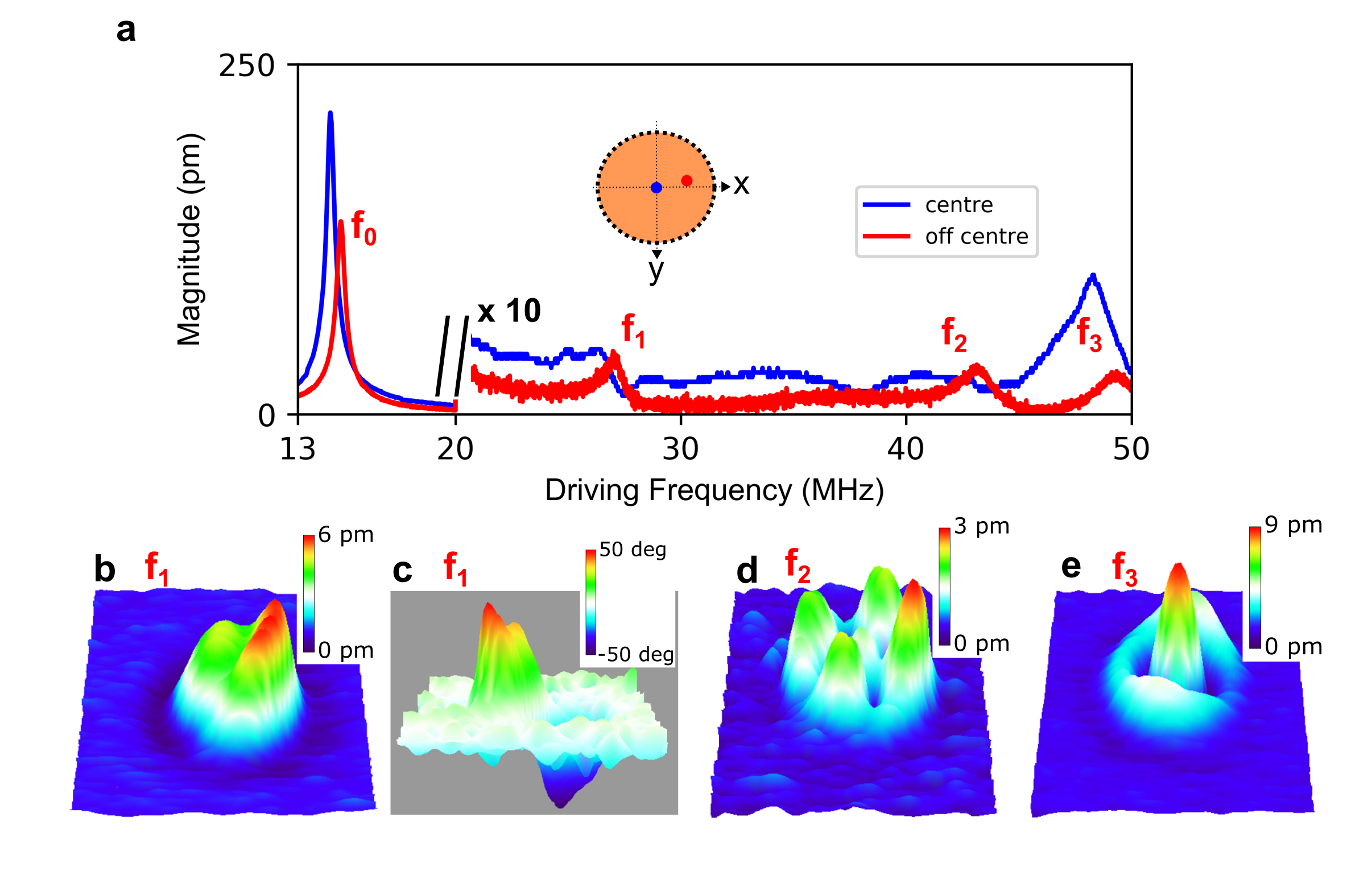}
    \caption{Resonant mode mapping of device A. (a) Response spectra retrieved at two different locations on the drumhead. For clarity, the magnitude of the spectra was scaled to 10x above 20 MHz of the driving frequency. The spectrum taken at the drum centre (shown in blue) has two pronounced peaks: the fundamental frequency at 14.46 MHz and a higher order resonance at 48.22 MHz. The off-centre spectrum (red) has four peaks labeled $f_0$ (at 14.55 MHz) through $f_3$. The inset shows location of the corresponding laser spot on the drumhead where the frequency response spectra are retrieved. (b)-(e) are all spatially resolved mapping measurements with the same x and y directions as in figure 1e and with a step size of approximately 189 nm in both x and y, with 64 steps in each direction. (b) Magnitude of $f_1$ (26.81 MHz) with z scale in pm. (c) The corresponding phase mapping of $f_1$ with scale in degrees is used to show that the two sections of the drum move in opposite directions. (d) Magnitude of $f_2$ (43.25 MHz) with z scale in pm. (e) Magnitude of $f_3$ (48.83 MHz) with z scale in pm.}
    \label{fig:Figure_2}
\end{figure}

\subsection{\label{sec:level2}Modal analysis}
\subsubsection{\label{sec:level3}Modal weights from experiment}
Figures 3a-e show the experimental off-resonance shapes $Z(x,y,\omega_d )$ of the circular drum, where $\frac{\omega_d }{2\pi }$ is between $f_2$ ((2,1) mode) and $f_3$ ((0,2) mode). From these, we can observe a transition of $Z(x,y,\omega_d )$ from $Z_{21} (x,y)$ to $Z_{02} (x,y)$. As $Z(x,y,\omega_d )$ evolves from one mapped mode shape to a neighbouring one, a natural thought would be to describe $Z(x,y,\omega_d )$ as a superposition of $Z_{mn} (x,y)$:
\begin{equation}
\mathbb{Z}(x,y,\omega_d )=\sum_{mn} \mathbb{Z}_{mn} (x,y)e_{mn} (\omega_d),     
\label{eq:Equation_1}
\end{equation}
where $e_{mn} (\omega_d)$ is the frequency-dependent weights of the corresponding resonance mode, $Z_{mn}$. Note that both $\mathbb{Z}(x,y,\omega_d )$ and $\mathbb{Z}_{mn} (x,y)$ are normalised so that $\int_{A}^{}\left |\mathbb{Z}(x,y)\right |^2 dxdy=1$ for the drum area $A$. Given $\mathbb{Z}(x,y,\omega_d)$ and $\mathbb{Z}_{mn} (x,y)$, $e_{mn} (\omega_d)$ is determined by the following integral:
\begin{equation}
e_{mn} (\omega_d )=\int_{A}^{} \mathbb{Z}_{mn}^* (x,y)\mathbb{Z}(x,y,\omega_d )dxdy,
\label{eq:Equation_2}
\end{equation}

The circle and square symbols from Fig. 3k show the dependence of the experimental modal weight, $\left |e_{mn} (\omega_d )\right |^2$, on the drive frequency, in percentage for the (2,1) and (0,2) modes. It is important to note here that the contributions of the other modes are not shown because their modal weights are almost zero (detailed results for all six modes are included in Supplementary Information). A similar analysis was also performed for the elliptical drum and the off-resonance shapes and corresponding experimental modal weights are shown in Figs. 4a-e and circle and square symbols in Fig. 4k, respectively. In both drums, a gradual transition of the modal weight dominance from (2,1) to (0,2) is observed. To confirm this trend, we used finite element method (FEM) simulations.

\subsubsection{\label{sec:level3}Modal analysis without damping using FEM simulation}

Figures 3f-j present the corresponding mechanical displacement $Z(x,y,\omega_d )$ obtained from FEM simulations using COMSOL neglecting damping parameters. The dotted lines in Fig. 3k are the simulation results and show a reasonable agreement with the experimental data. We see that in the absence of damping, participation at the resonance frequencies is equal to 100\%, which is not always the case in the experiment. From here we can see the transition of the  modal weight dominance from (2,1) mode to (0,2). The FEM results show that at 44 MHz, the contributions are ~45\% and ~55\% for the (2,1), and (0,2) modes, respectively. In contrast, at 48 MHz, the response shape is overwhelmingly dominated by the (0,2) mode, while (2,1) mode contributes ~0.07\% only. 

\begin{figure}
    \centering
    \includegraphics[width=1.0\columnwidth]{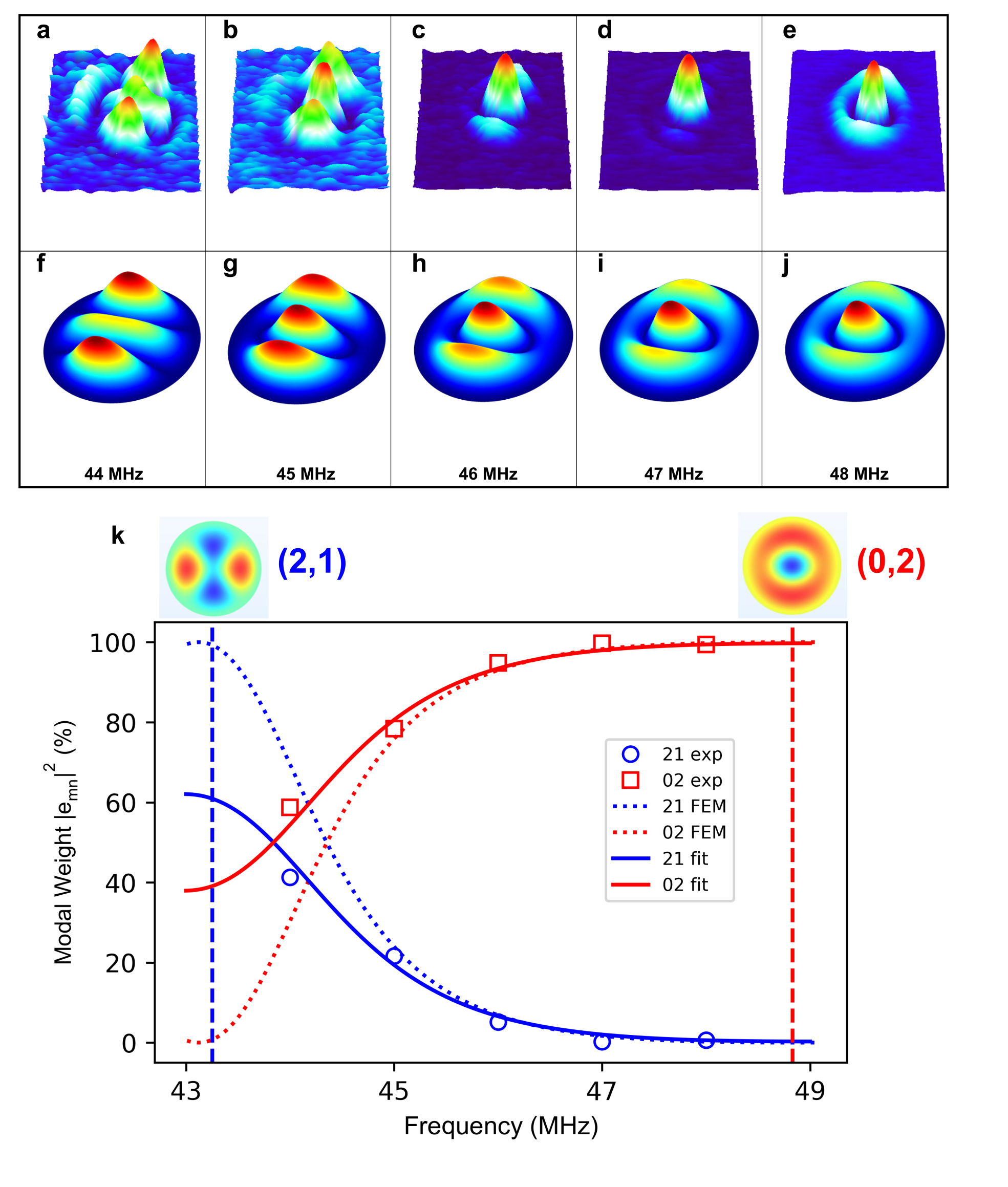}
    \caption{Off-resonance modes of device A. (a)-(e) Magnitude mapping at driving frequencies between the (2,1) mode through the (0,2) mode. (f)-(j) Corresponding simulated magnitude mapping. All mappings have the same (x,y) directions as in figure 1e. (k) Modal weight for (2,1) and (0,2) modes as a function of the driving frequency: the circle and square symbols are obtained from the experiment using Eq. (2), the dotted lines are from the FEM simulations, and the solid lines represent the fitting using Eq. (6). The vertical dashed lines indicate the resonance frequencies of the (2,1) and (0,2) modes with the corresponding modal shapes shown above.}
    \label{fig:Figure_3}
\end{figure}

We have done the same analysis for the elliptical drum. Figures 4f-j shows the simulated $Z(x,y,\omega_d )$ between (2,1) and (0,2) modes and the corresponding modal weights are shown as dotted lines in Fig. 4k. Similar to the circular drum, the transition of modal weight dominance from (2,1) to (0,2) can also be observed.

\begin{figure}
    \centering
    \includegraphics[width=1.0\columnwidth]{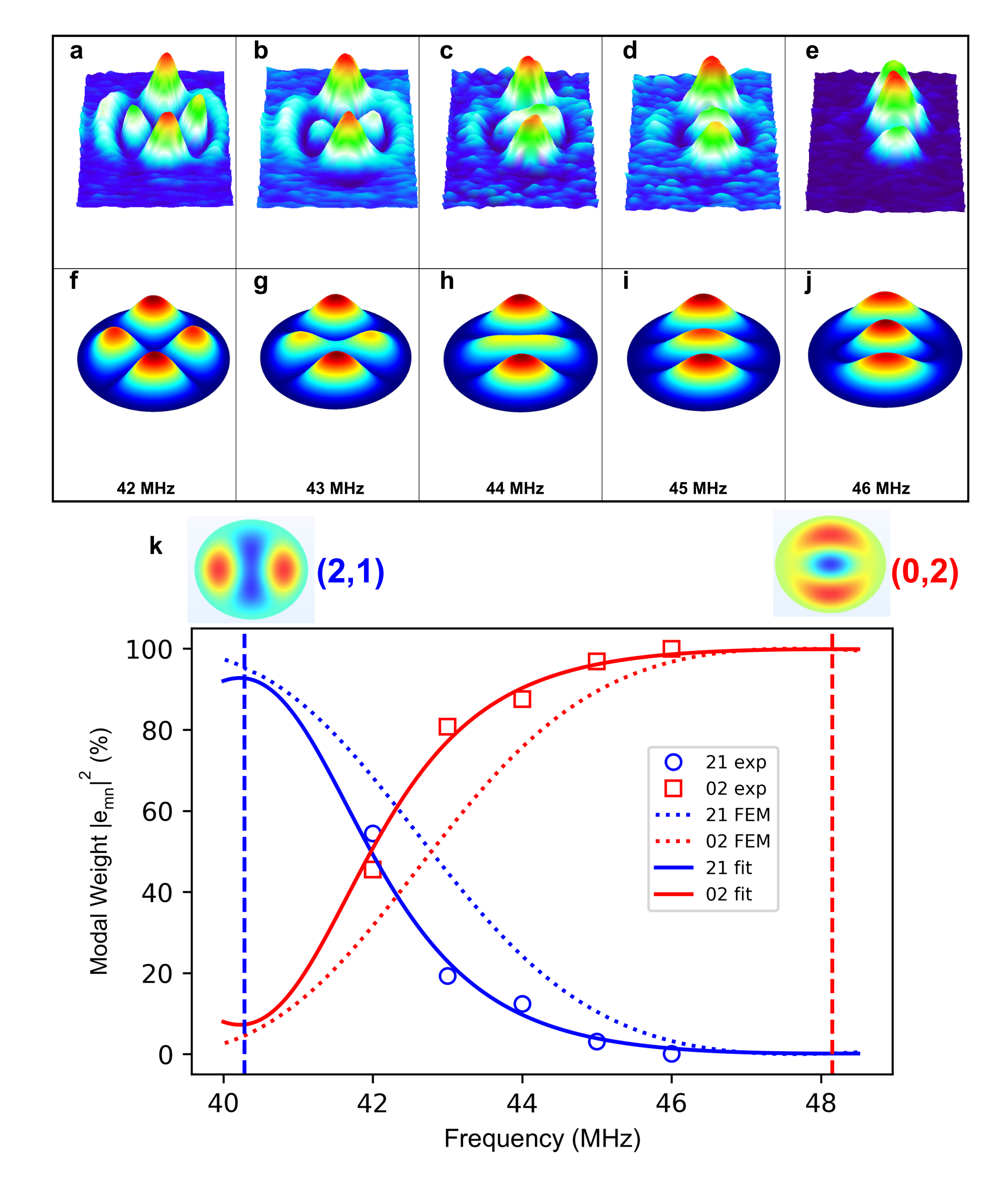}
    \caption{Off-resonance modes of device B. (a)-(e) Magnitude mapping at driving frequencies between the (2,1) mode through the (0,2) mode. (f)-(j) Corresponding simulated magnitude mapping. All mappings have the same (x,y) directions as in figure 1e. (k) Modal weight for (2,1) and (0,2) modes as a function of the driving frequency: the circle and square symbols are obtained from the experiment using Eq. (2), the dotted lines are from the FEM simulations, and the solid lines represent the fitting using Eq. (6). The vertical dashed lines indicate the resonance frequencies of the (2,1) and (0,2) modes with the corresponding modal shapes shown above.}
    \label{fig:Figure_4}
\end{figure}

\subsubsection{\label{sec:level3}Modal weight dependence on driving frequency}
In the observations presented above, we are only able to infer the trend of $e_{mn} (\omega_d)$. To better understand the frequency dependence of $e_{mn} (\omega_d)$, we start with the general equation of motion for a forced clamped plate\cite{Timoshenko1928}:
\begin{equation}
\rho h\frac{\partial^2z_{dc+ac}(x,y,t)}{\partial t^2} + D\nabla^4z_{dc+ac}(x,y,t) = \frac{\varepsilon_{0}(V_{dc}+V_{ac}\cos(\omega t))^2}{2(g_{0}-z_{dc+ac}(x,y,t))^2},
\label{eq:Equation_3}
\end{equation}
where $z_{dc+ac}(x,y,t)$ is the out-of-plane displacement, whose Fourier transform is $Z(x,y,\omega_d)$, $\rho$ is the mass density of the plate material, $h$ is the plate thickness, and $D$ is the flexural rigidity. The right-hand side of equation 3 is the total driving force per unit area\cite{davidovikj2017,Sajadi2017} $F_{dc+ac}(x,y,t)$, where $\varepsilon_0$ is the vacuum permittivity, and $g_0$ is the distance between the drumhead and the bottom electrode. Under the applied force, the total displacement can be presented as a sum of the static and oscillating displacements: $z_{dc+ac}(x,y,t) = z_{dc}(x,y) + z_{ac}(x,y,t)$. The interferometric detection employed in this work allows the observation of the modal responses to the AC drive.  Therefore, we simplify equation 3 to focus on the AC response, $z_{ac}$, to the AC component of the driving force, $F_{ac}$. Furthermore, since $V_{dc} \gg V_{ac}$, $F_{ac}$ can be further simplified by assuming $z_{dc+ac}$ to be roughly equal to $z_{dc}$ (see Supplementary Information Note 2 for the derivation details). The results of these simplifications will give the following equation of motion\cite{davidovikj2017,huang2018}:
\begin{equation}
\rho h\frac{\partial^2z_{ac}(x,y,t)}{\partial t^2} + D\nabla^4z_{ac}(x,y,t) - \frac{\varepsilon_{0}V_{dc}^2}{(g_{0}-z_{dc}(x,y))^3}z_{ac}(x,y,t)
=  \frac{\varepsilon_{0}V_{dc}V_{ac}\cos(\omega t)}{(g_{0}-z_{dc}(x,y,t))^2}
\label{eq:Equation_4}.
\end{equation}
The right-hand side of equation 4 is now the AC component of the driving force per unit area, $F_{ac}(x,y,t)$.
\\\indent
The next step is to approximate the shape of $z_{dc}$. For a clamped circular or elliptical plate under a uniform load, the shape of the deformation can be described by the Bessel function for the (0,1) mode with the following form\cite{krauthammer2001,Wong2010,kelly2013,sajadi2018,korenev2002}:
\begin{equation}
z_{dc}(x,y) = Bg_{0}\left (1-\left (\frac{x^2}{a^2}+\frac{y^2}{b^2}\right )\right )^2
\label{eq:Equation_5},
\end{equation}
where $a$ and $b$ are the motional semi-major and semi-minor axes for the elliptical drum. In the case of the circular drum, $a=b=R$, which is the motional radius of the circular drum. The parameter $B$ here is the static deformation prefactor that reflects the sharpness of the distribution. Finally, using equations 1 and 5, $e_{mn} (\omega_d )$ takes the following form (see Supplementary Information Note 3 for the derivation details):
\begin{equation}
e_{mn}(\omega_d) = \frac{\int_{A}^{}\mathbb{Z}_{mn}(x,y)F_{ac}(x,y)dxdy}{\omega_{mn}^2-\omega_{d}^2+i\gamma_{mn}\omega_d},
\label{eq:Equation_6}
\end{equation}
where $\gamma_{mn}$ is a phenomenological damping parameter for ($m,n$) mode, and $F_{ac} (x,y)$ is the spatial distribution of the AC driving force amplitude. The time-dependent driving force has the form: $F_{ac} (x,y,t) = F_{ac} (x,y)\cos(\omega t)$. We take the experimental values of the damping parameters from their aforementioned $Q$ factors, which are 2.39 MHz and 2.35 MHz for $\gamma_{21}$ and $\gamma_{02}$ of the circular drum, and 1.30 MHz and 1.99 MHz for $\gamma_{21}$ and $\gamma_{02}$ of the elliptical drum, and use $B$ as a fitting parameter in equation 6. The results of the fitting shown in Figs. 3 and 4, as previously mentioned, yields $B = 0.12$ for the circular drum and $B = 0.32$ for the elliptical drum. Figure 5 shows how the AC force spatial distribution looks like for the drums given their respective values of $B$. The vertical axis is $F_{ac}(x,y)/|F_{ac}|_{edge}$, where $|F_{ac}|_{edge} = \varepsilon_{0}V_{dc}V_{ac}/g_0$. For the circular drum, the force spatial distribution looks almost uniform across the drumhead. On the other hand, the distribution in the elliptical drum is sharper compared to the circular drum. This difference in sharpness could be explained by the difference in the drums’ geometry. To support this, we look at the ratio $\frac{B_{ellipse}}{B_{circle}}$ from the experiment and compare it to the theoretically derived expression\cite{krauthammer2001} $\frac{\beta_{ellipse}b^4}{\beta_{circle}R^4}$, where $\beta$ is an eccentricity scaling factor. The ratio from the experimental fit yields 2.67 and is close to 2.54, which is derived from the theoretical expression using $b$, $R$ and $\beta$ from the reference\cite{krauthammer2001}. This difference in sharpness implies that the electrostatic deformation of the elliptical drum is slightly bigger than that of the circular drum due to the geometric difference\cite{Chen2013,Landau2013}. In other words, a slight difference in static deformation translates into a sharper difference in the driving force as $F_{ac}\sim1/z_{dc}^2$.

\section{Discussion}
\subsection{\label{sec:level2}Effect of driving force on the modal shapes}
Equation 6 allows us to infer two significant insights. First, the frequency dependence of the modal weight $e_{mn} (\omega_d )$ is a Lorentzian, which can be interpreted as the frequency response of a damped harmonic oscillator with a resonance frequency $\omega_{mn}$. Second, $e_{mn} (\omega_d )$ reflects the relation between the force distribution and shape of the ($m,n$) mode. This matching is quantified by $\int_{A}^{}\mathbb{Z}_{mn}(x,y)F_{ac}(x,y)dxdy$. The force distribution can influence the participation of a particular resonance mode. In both drums, the force distribution is axisymmetric, since the bottom electrode is global. Because of this, the force distribution accentuates the axisymmetric modes such as (0,2) and diminish the participation of non-axisymmetric modes such as (2,1). This is the reason why, the modal equilibrium points, the points at which the two modes have equal weights, for both drums are closer to the (2,1) mode. In other words, the dominance of the (0,2) mode can be observed in a broader frequency range compared to the (2,1) mode.
\\\indent
From the point of view of drum geometry, we compared the modal equilibrium points of the drums with each other. We see that the equilibrium is closer to the (2,1) mode in the circular drum than the elliptical drum. This means that the additional eccentricity present in the elliptical drum allows the (2,1) mode to widen its dominance in frequency\footnote{The details of the effect of eccentricity on the modal equilibrium position is shown in Supplementary Information}. In fact, the presence of the (2,1) mode implies that there is an inherent asymmetry in the system. A perfect circular drum would only show axisymmetric modes.
\\ \indent
\begin{figure}
    \centering
    \includegraphics[width=1.0\columnwidth]{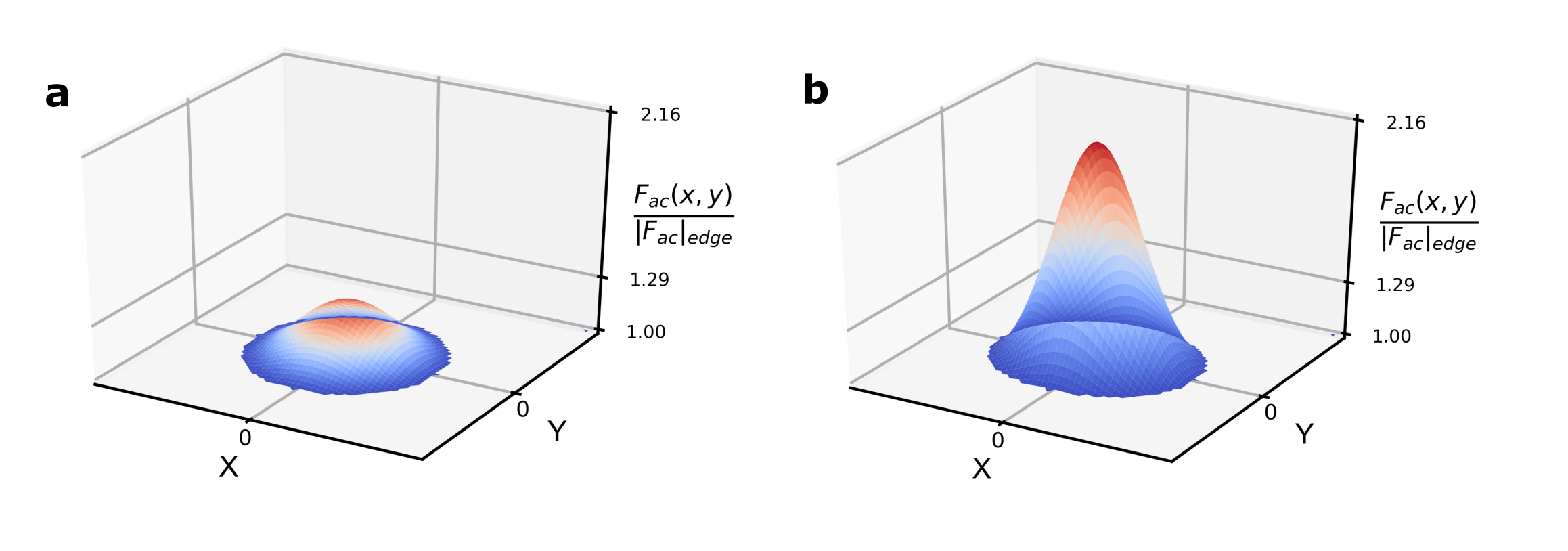}
    \caption{Simulated AC driving force spatial distribution. (a) and (b) illustrate the driving force spatial distribution for the circular and elliptical drum, respectively. $B_{circle} = 0.12$, which gives maximum $F_{ac}(x,y)/|F_{ac}|_{edge} = 1.29$, and $B_{ellipse} = 0.32$, which gives maximum $F_{ac}(x,y)/|F_{ac}|_{edge} = 2.16$}
    \label{fig:Figure_5}
\end{figure}
 This type of analysis can be relevant to coupling of NMRs to other systems\cite{LeeJ2018}. Firstly, for intermodal coupling, while it is typical to utilize the fundamental mode, there might be cases where the higher modes are more appropriate to use. One of these works include using multiple electrodes to enhance the driving of the non-axisymmetric modes and show tunable intermodal coupling\cite{DeAlba2016}. Another used parametric excitation to enhance the intermodal coupling\cite{mathew2016}. Furthermore, 2D NMRs' resonant modes can couple to multiple external fields. For example, through the radiation-pressure force\cite{aspelmeyer2014}, NMRs have been found as efficient, bidirectional, and coherent transducers between the microwave and optical regions of the electromagnetic spectrum\cite{Andrews2014}. The large quantum point motion of 2D NMRs stimulated the growth of numerous studies that couple mechanical resonators to superconducting cavities\cite{singh2014,noguchi2016}. By carefully choosing 2D materials with some exotic properties, such as a superconducting NbSe$_2$\cite{sengupta2010,will2017}, a hybrid system of microwave and optical subsystems mediated by 2D NMRs could be realised. Finally and more importantly, this type of analysis is universal. For example, even with a different geometry, the global distribution of the magnetomotive driving force explains the favourability of the detection of the odd modes for the single beam resonators\cite{westra2010}$^,$\footnote{FEM simulations of a beam geometry with global driving force distribution is shown in Supplementary Information}. The wealth of potential studies involving coupled NMRs makes the knowledge gained from the effect of the  driving force on the resonance modes an invaluable tool.
\\ \indent
In summary, we observed intermodal transitions of NbSe$_2$ mechanical plate drums through spatial mapping at off-resonance frequencies. We, then, described the off-resonance motion as a superposition of the resonance modes. Through this modal analysis, we were able to see how participation of resonance modes changes as the driving frequency changes. Furthermore, by looking at the modal weight formula, we were able to describe how the modal shapes are revealed through the application of the driving force and deduce how this driving force is distributed across the oscillator. In fact, all mechanical vibrations of objects are mixtures of their resonance modes across the driving frequency spectrum with varying weights of participation. This fundamental and universal understanding of the relationship between the resonance modes and the driving force will greatly benefit all future NMRs studies, which will inevitably involve coupling of the flexural motion to various degrees of freedom of different nature and energy.

\noindent{\small \textbf{\textsf{METHODS}}}
\\\indent
\textbf{Sample Fabrication.}
40 nm Au and 20 nm Cr electrodes were lithographically patterned on 7 mm x 7 mm x 0.65 mm p-doped Si chips with a thermally grown 543 nm thick SiO$_2$ layer. The chip is then cleaned through ultrasonication for 10 minutes in acetone, 2 minutes in IPA, and 1 minute in DI water. The drums' spacer is then created by spin coating the chip with the AR-P (CSAR-62) electron beam resist. After baking at 180$^o$C for one minute, the resist is patterned with the drum hole and contact window patterns. After development, the resist is baked again at 180$^o$C for 9 minutes to make it rigid. The spacer thickness is measured to be 295$\pm$10 nm using a commercial stylus profilometer. Bulk NbSe$_2$ purchased from HQ Graphene are exfoliated and transferred onto the patterned drum and contact windows using a deterministic dry PDMS stamp transfer process\cite{castellanos2014,pande2020}. 
\\\\
\noindent{\small \textbf{\textsf{AUTHOR CONTRIBUTIONS}}}
\\\indent
C.D.C. conceived the device and supervised the project; J.C.E. fabricated the samples; K.-H. L. and C.-Y. Y. designed and built the setup for optical measurements; J.C.E., M.A.C.A., and C.-Y.Y performed the measurements; J.C.E., M.A.C.A., J.Y.W., S.K., Y.P., and C.D.C analyzed the data, performed simulations, and wrote the manuscript; all authors discussed the results and contributed to the manuscript.
\\\\
\noindent{\small \textbf{\textsf{CONFLICT OF INTEREST}}}
\\\indent
The authors declare no conflict of interest.

\begin{acknowledgments}
We acknowledge the contributions of Tzu-Hui Hsu and Wen-Hao Chang in the fabrication of the devices and building the experimental setup. We also thank the Taiwan International Graduate Program for the financial support. This project is funded by  Academia Sinica Grand Challenge Seed Program (AS-GC-109-08), Ministry of Science and Technology (MOST) of Taiwan (107-2112-M-001-001-MY3), Cost Share Programme (107-2911-I-001-511), the Royal Society International Exchanges Scheme (grant IES$\backslash$R3$\backslash$170029), and iMATE (2391-107-3001). We would also like to extend our gratitude for the Academia Sinica Nanocore facility.
\end{acknowledgments}

%aipnum4-2.bst 2019-01-14 (MD) hand-edited version of apsrev4-1.bst
%Control: key (0)
%Control: author (8) initials jnrlst
%Control: editor formatted (1) identically to author
%Control: production of article title (0) allowed
%Control: page (1) range
%Control: year (1) truncated
%Control: production of eprint (0) enabled
%

\end{document}